\documentclass[twocolumn,prb,superscriptaddress]{revtex4}
\usepackage{graphicx}
\usepackage{dcolumn}% Align table columns on decimal point
\usepackage{bm}% bold math
\usepackage{epstopdf}
\makeatletter

\newcommand{\Rmnum}[1]{\expandafter\@slowromancap\romannumeral #1@}
\makeatother
\usepackage[usenames]{color}
\begin{document}

\title{Coexistence of high electrical conductivity and weak ferromagnetism in $Cr$ doped $Y_2Ir_2O_7$ pyrochlore iridate}
\author{Vinod Kumar Dwivedi}
\email{vinodd@iitk.ac.in}
\affiliation{Materials Science Programme, Indian Institute of Technology, Kanpur 208016, India}
\author{Soumik Mukhopadhyay}
\email{soumikm@iitk.ac.in}
\affiliation{Department of Physics, Indian Institute of Technology, Kanpur 208016, India}
%\date{\today}
\begin{abstract}
We report the structural, magnetic and electrical transport properties of $Y_2Ir_{2-x}Cr_xO_7$ pyrochlore iridates. The chemical doping leads to order of magnitude enhancement of electrical conductivity. The introduction of $Cr^{3+}$ at $Ir^{4+}$-site tends to distort the $Ir-O_6$ octahedra and suppresses antiferromagnetic correlation. The X-ray photoemission spectroscopy measurements suggest the coexistence of $Ir^{4+}$ and $Ir^{5+}$ valence states in the $Y_2Ir_{2-x}Cr_xO_7$ compounds. The concentration of $Ir^{5+}$ is enhanced with $Cr$ doping, leading to weak ferromagnetism and enhanced electrical conductivity. A cluster-glass like transition is also observed at low temperature with $Cr$ doping, possibly due to competing ferromagnetic and antiferromagnetic interaction.
\end{abstract}
%\pacs{71.27.+a, 78.70.Dm, 72.20.-i, 72.15.Rn, 78.20.Ek}

\maketitle
\section{Introduction}
The interplay of electron correlation, spin-orbit coupling (SOC), crystal field effect and geometric frustration can lead to many emergent quantum phases and interesting phenomenology such as spin-liquid ~\cite{Machida}, spin-ice~\cite{Ramirez}, spin glass~\cite{Aito,Soda,Gingras,Yoshii,Harish}, anomalous Hall effect~\cite{Machida1}, frustrated Kondo lattice~\cite{Machida}, superconductivity~\cite{Hanawa}, etc. in 5d transition metal oxides in general and pyrochlore iridates $A_2Ir_2O_7$ in particular~\cite{Wan,Pesin,Yang,Hongbin,Krempa,Abhishek1,Abhishek2}. For $Y_2Ir_2O_7$, the magnetic properties are determined by the contribution from $Ir^{4+}$ ion and the complex magnetic ground states emerging from $f-d$ exchange interactions are avoided~\cite{Gang}. This allows us to study Ir order by separating out the properties emerging from interaction between the rare-earth and $Ir^{4+}$ ions. $Y_2Ir_2O_7$ is expected to be a Weyl semimetal with all-in/all-out antiferromagnetic (AFM) ground state~\cite{Wan}. Since the ground state of pyrochlore iridates are sensitive to the SOC, crystal field effect and on-site Coulomb repulsion (governed \textit{inter-alia} by the $Ir-O-Ir$ bond angle and $Ir-O$ bond length), a small change in A site radius by chemical doping or substitution at the Ir site may easily alter the balance between the competing energies with varying consequences~\cite{Tafti,Wan1,Koo,Harish2,Harish3,Yanagishima,Matsuhira}.

 The low temperature magnetic state of $Y_2Ir_2O_7$ is still debated, because neutron diffraction and inelastic scattering analysis do not show clear evidence of long range magnetic ordering~\cite{Shapiro,Disseler1}. On the other hand, muon spin rotation and relaxation experiment analysed with \textit{ab} \textit{initio} modeling~\cite{Disseler2} suggest a long range magnetic transition with all-in/all-out AFM ground state at low temperatures. There are few earlier reports which also suggest existence of weak ferromagnetic component over the AFM background in $Y_2Ir_2O_7$~\cite{Zhu,Vinod1,Vinod2}. Since, pyrochlore iridates are geometrically frustrated, therefore the low temperature magnetic state of $Y_2Ir_2O_7$ could show glass-like behaviour~\cite{Aito,Harish,Fukazawa,Taira}, similar to other Y-based pyrochlores such as $R_2Os_2O_7$ (R = Y and Ho)~\cite{Zhao}, $Y_2Ru_2O_7$~\cite{Yoshii} and $Y_2Mo_2O_7$~\cite{Gingras}. Despite these advances, a conclusive understanding of the precise nature of the low temperature magnetic state is far from being realized. While hole doping by the introduction of $Ca^{2+}$ at $Y^{3+}$ site in $Y_2Ir_2O_7$ compound modifies the Ir-$t_{2g}$ electron band width leading to enhanced electrical conductivity and ferromagnetism~\cite{Fukazawa,Zhu}, replacement of $Y^{3+}$ with isovalent ion $Bi^{3+}$, on the other hand, also shows insulator to metal transition and spin-glass like magnetic transition~\cite{Aito,Soda}, similar to $Y_{2-x}Bi_xRu_2O_7$~\cite{Yoshii} and $Bi_{2-x}Y_xRu_2O_7$ compounds~\cite{Kanno}. In the context of electrical charge transport in nanoscale systems, a gapped out Weyl-Semimetal phase has also been reported in $Y_{2-x}Bi_xIr_2O_7$~\cite{Abhishek1,Abhishek2}. Previous studies on doping of isovalent magnetic $Ru^{4+} (4d^4)$~\cite{Harish2} and non-magnetic $Ti^{4+} (3d^0)$ ~\cite{Harish3} ions separately at magnetic $Ir^{4+}$-site in $Y_2Ir_2O_7$ compounds show enhancement in AFM correlation with negligible increase in electronic conductivity. However, the influence of substitution of magnetic $3d$ ions at magnetic $Ir^{4+} (5d^5)$-site in $Y_2Ir_2O_7$ has not been investigated yet. 

 In the present work, we have studied the effect of magnetic $Cr$ ion introduction at magnetic $Ir$-site in $Y_2Ir_{2-x}Cr_xO_7$. Since, the $Ir^{4+} (5d^5)$ exhibit stronger SOC and less on-site Coulomb repulsion (U) as compared to $Cr^{3+} (3d^3)$, the doping of $Cr^{3+}$ at $Ir^{4+}$ site acts not only as hole doping, but at the same time likely to reduce the SOC and amplify U. We show that $Cr$ doping at magnetic $Ir^{4+}$-site in $Y_2Ir_2O_7$ compound leads to cluster-glass like behaviour and orders of magnitude enhancement of electrical conductivity with interesting consequences on the transport and magnetic properties.

\section{Experimental Methods}
The parent sample $Y_2Ir_2O_7$ was prepared by solid state reaction method reported by same authors~\cite{Vinod1,Vinod3,Vinod4}. On the other hand, $Y_2Ir_{2-x}Cr_xO_7$, $x = 0.05,0.1,0.2$ series were prepared using the method described elsewhere~\cite{Aito,Soda}. The crystal structure was analyzed by powder X-ray diffraction (XRD) using a PANalytical XPertPRO diffractometer with Cu K$\alpha$ radiation ($\lambda$= 1.54 $\AA$) at room temperature. The actual composition of the samples were determined using energy dispersive x-ray spectrometry (EDX) with the help of field emission scanning electron microscope (FE-SEM) [JSM-7100F, JEOL]. Electrical transport properties were measured by conventional four probe technique. The magnetization measurements were performed in a Quantum design physical property measurement system (PPMS). The electronic structure was characterized by the X-ray photoelectron spectroscopy (XPS) using a PHI 5000 Versa Probe II system.

\section{Results and Discussion}
\begin{figure}
\includegraphics[width=9cm]{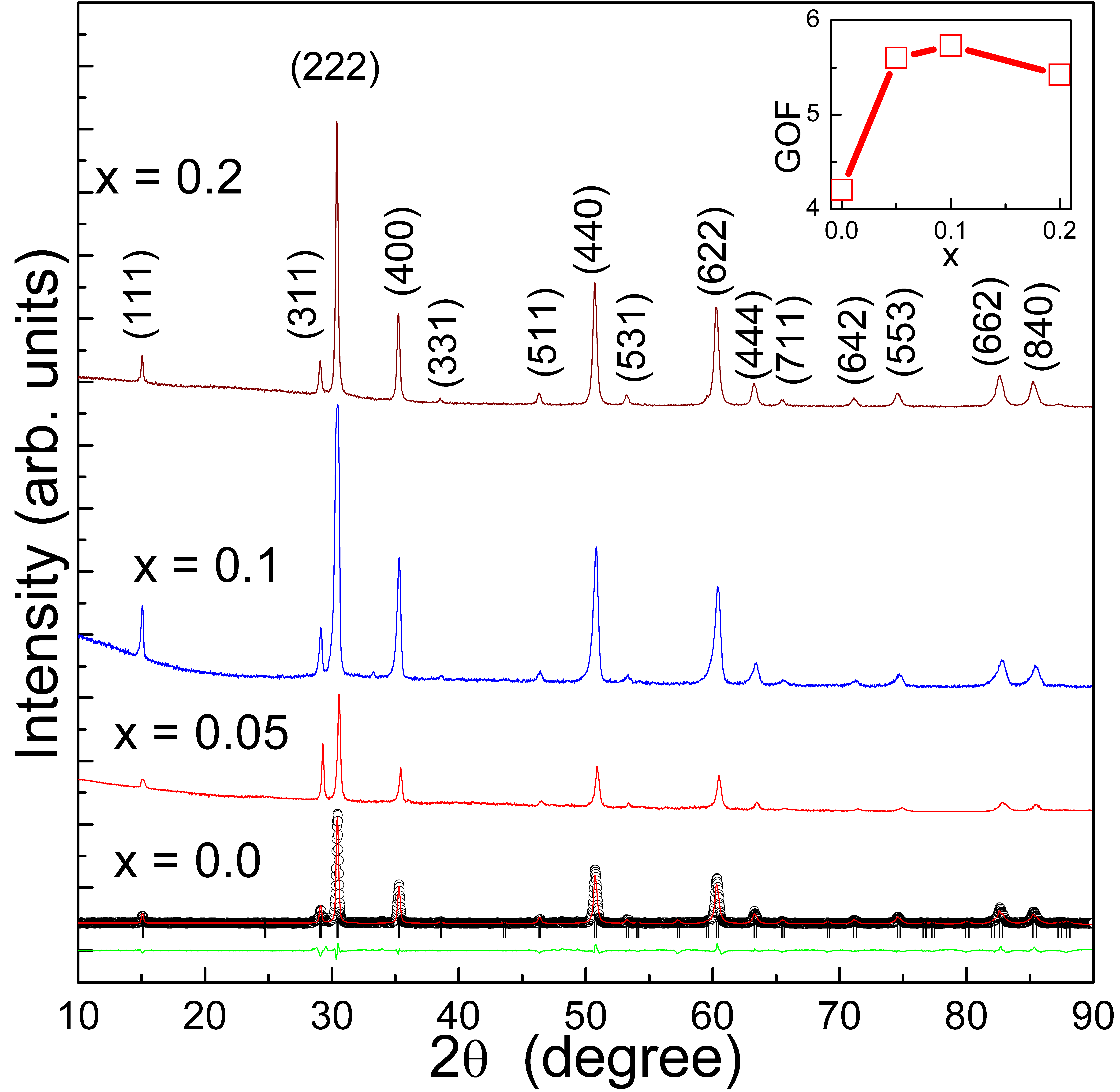}\\
\caption{Refined XRD profile of $Y_2Ir_{2-x}Cr_xO_7$: Inset shows goodness of fit.}\label{fig:xrd}
\end{figure}

\begin{figure}
\includegraphics[width=8.5cm]{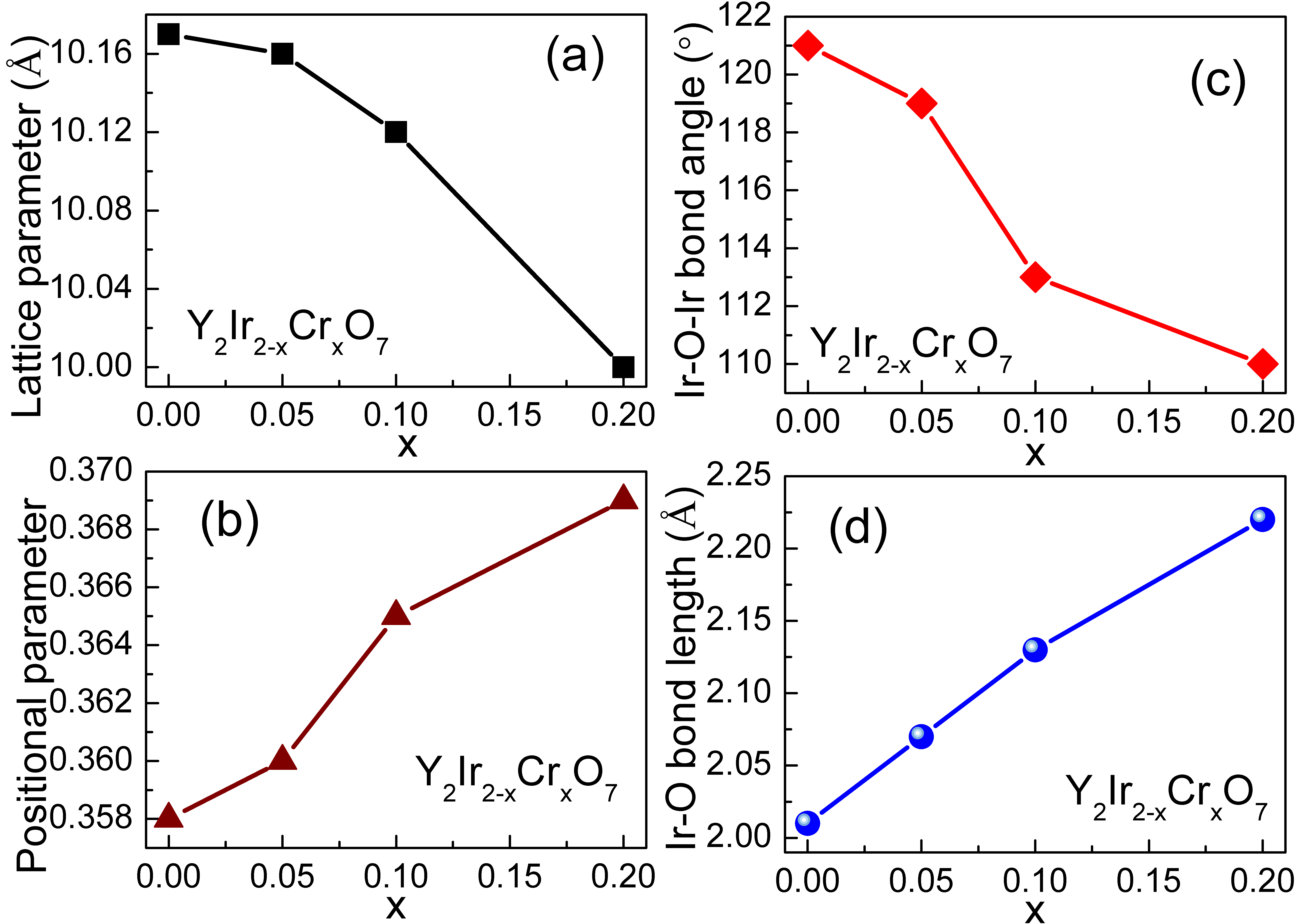}\\
\caption{(a) Lattice parameter a, (b) positional parameters $p$ of $O1$ atom, (c) $Ir-O-Ir$ bond angle, and (d) $Ir-O$ bond length as a function of Cr doping concentration x, of the sample $Y_2Ir_{2-x}Cr_xO_7$.}\label{fig:bond}
\end{figure}

Figure~\ref{fig:xrd} shows the powder XRD patterns along with Rietveld refinement for $Y_2Ir_{2-x}Cr_xO_7$ samples at room temperature. Inset of Fig.~\ref{fig:xrd} shows goodness of fit (GOF) defined as $F^2 = [R_{wp}/R_{exp}]^2$, where, $R_{wp}$ is the expected weighed profile factor and $R_{exp}$ the observed weighed profile factor~\cite{Rietveld1}. The accuracy of all refined values of XRD data is $\sim \pm 0.002^0$. Analysis of XRD spectra indicate that all the samples crystallize in the F-centered cubic unit cell with $Fd\bar{3}m$ symmetry, which is consistent with previous reports~\cite{Zhu,Vinod1}. XRD pattern shows no major changes in the peak positions with doping. Note the slight mismatch in ionic radii between $Ir^{4+}$ = 0.625 $\AA$ and $Cr^{3+}$ = 0.615 $\AA$~\cite{Shannon}. Interestingly, Fig.~\ref{fig:bond}a shows that the lattice constant decreases marginally for $Y_2Ir_{2-x}Cr_xO_7$ (YICO). The cubic pyrochlore oxide with general formula of $A_2Ir_2(O1_6)(O2)$ having space group Fd$\bar{3}$m has 8 atoms per unit cell. The four non equivalent atoms occupy the following positions: $A$ at $16d$ site ($\frac{1}{2}$, $\frac{1}{2}$, $\frac{1}{2}$), $Ir$ at $16c$ site (0, 0, 0), $O1$ at $48f$ site ($p$, $\frac{1}{8}$, $\frac{1}{8}$) and $O2$ at $8b$ site ($\frac{3}{8}$, $\frac{3}{8}$, $\frac{3}{8}$), with $p$ being the only one adjustable positional parameter. In this pyrochlore structure $16d$ sites contain larger $A$-type cations thus forming an axially compressed scalenohedron which are coordinated to six $O1$ atoms and two $O2$ atoms. The $16c$ site exhibit smaller $Ir$-cation coordinated with six $O1$ atoms at equal distances from the central $Ir$ ion in a trigonal antiprism. The shorter $A-O2$ bond length depends only on lattice constant, while the $A-O1$ and $Ir-O2$ distances depend on both lattice parameters and the position parameters. For $p = 0.3125$ one has a perfect octahedron about $16c$ site where $Ir$ cations reside under a perfect cubic field~\cite{Gardner}. Figure~\ref{fig:bond}b shows the positional parameter $p$ as a function of doping content x. The value of $p$ for un-doped sample turns out to be 0.355, larger than the ideal value suggesting compressed and distorted octahedra of $IrO_6$. $p$ value increases as $Cr$ concentration increases in $Y_2Ir_{2-x}Cr_xO_7$ samples suggesting more distorted and elongated $IrO_6$ octahedra, which gives rise to enhanced crystal fields. The doping dependence of $Ir-O-Ir$ bond angle and $Ir-O$ bond length are shown in Fig.~\ref{fig:bond}c, and d, respectively. The bond angle decreases and bond length increases with increasing $Cr$ content in $Y_2Ir_{2-x}Cr_xO_7$ samples. This results in more distortion in $IrO_6$ octahedra and enhancement in mixing between $Ir$($5d$)/$Cr$($3d$) and $O$($2p$), $Y$($4p$) states.

\begin{figure}
\includegraphics[width=9cm]{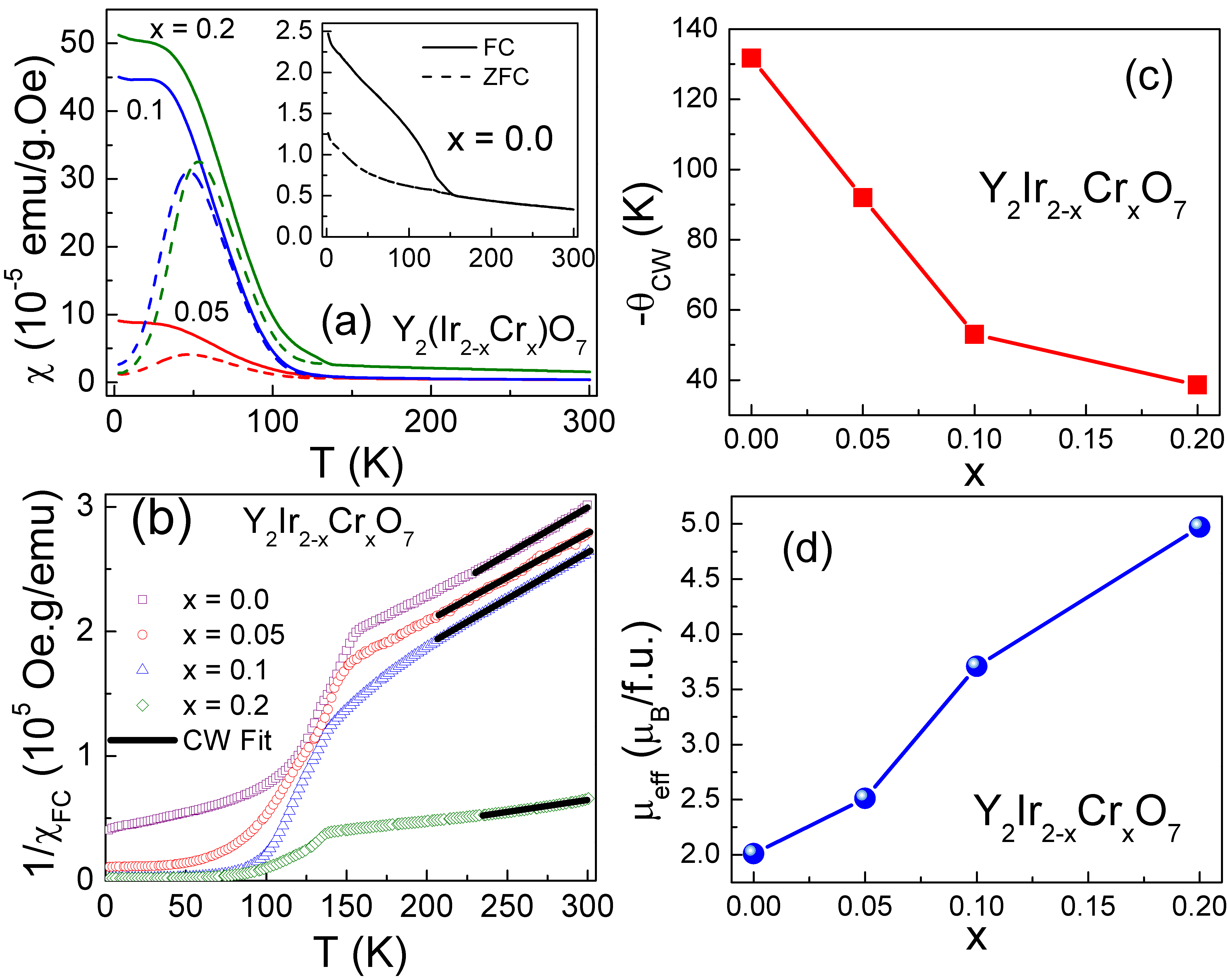}\\
\caption{(a) Magnetic susceptibility ($\chi$ = M/H) as a function of temperature (T) measured in 1000 Oe applied magnetic field following the protocol of zero field cooled (ZFC) and field cooled (FC) for x = 0.05, 0.1 and 0.2 samples: inset shows $\chi (T)$ of x = 0.0 sample. (b) Inverse of magnetic susceptibility $\chi^{-1}$ as a function of temperature (T) for x = 0.0, 0.05, 0.1 and 0.2 samples. Red dashed line displays the fit using the Curie-Weiss (CW) law. (c) Curie-Weiss temperature $\theta_{CW}$, and (d) Effective magnetic moment $\mu_{eff}$ against doping concentration.} \label{fig:mt}
\end{figure}

The temperature dependence of the zero field cooled (ZFC) and field cooled (FC) magnetic susceptibilities for $Y_2Ir_{2-x}Cr_xO_7$ (x = 0.05, 0.1 and 0.2) is shown in Fig.~\ref{fig:mt}a. For undoped (x = 0.0) sample, irreversibility sets in between the FC and ZFC magnetization below $\sim$ 160 K [ inset of Fig.~\ref{fig:mt}a] suggesting a magnetic transition, consistent with previous reports~\cite{Zhu,Vinod1}. Recent studies seem to suggest coexistence of a weak FM component on the large AFM background~\cite{Zhu,Vinod1}. As shown in the Fig.~\ref{fig:mt}a, with $Cr$ doping the FC magnetic susceptibility is enhanced and the temperature for the magnetic irreversibility $T_{irr}$ shifts towards the lower temperature side. This might be due to the double exchange interaction between neighbouring $Ir^{4+}$, $Ir^{5+}$/ and $Cr^{3+}$ ions via $Ir^{4+}-O^{2-}-Ir^{5+}$ and $Ir^{4+}-O^{2-}-Cr^{3+}$ paths in terms of the magnetically compatible electronic orbitals $t_{2g}^3e_g^0$ between $Ir^{4+}$ and $Cr^{3+}$ ions. It can be noticed that ZFC magnetization for $Cr$ doped samples show distinct cusp at lower temperature compared to $T_{irr}$ [Fig.~\ref{fig:mt}a] suggesting a cluster-glass-like transition. This cusp is absent in the parent compound [inset of Fig.~\ref{fig:mt}a]. 

 At high temperature, the temperature dependence of susceptibility for all the samples is described by the Curie-Weiss (CW) law, $\chi = \frac{C}{T-\theta_{CW}}$, where $C$ and $\theta_{CW}$ are the Curie constant and Curie-Weiss temperature, respectively shown in Fig.~\ref{fig:mt}b. The estimated value of $\theta_{CW}$ for undoped compound show consistency with few reported values~\cite{Vinod1,Hui,Hui1}, however it shows large deviation with other reported values~\cite{Harish,Harish2,Harish3,Zhu}. This discripancy is due to the choice of fitting parameters, particularly the temperature independent constatnt term $\chi_0$ in $\chi = \frac{C}{T-\theta_{CW}} + \chi_0$. We have also fitted the data using this extra term $\chi_0$ and the estimated $\theta_{CW}$ turns out to be consistent with values reported in Reference~\cite{Harish,Harish2,Harish3}. The negative value of $\theta_{CW}$ for un-doped as well as doped samples suggests AFM correlation. The absolute value of $\theta_{CW}$ temperature decreases with increased $Cr$ doping [Fig.~\ref{fig:mt}c], which indicates the weakening of AFM coupling. It can be noticed that negative $\theta_{CW}$ decreases but the actual ordering temperature increases with $Cr$ doping content. This suggests reduction of frustration parameter [$f = \theta_{CW}/T_{irr}$] with $Cr$ doping, leading to large enhancement of the magnetization.

We have calculated the effective magnetic moment $\mu_{eff}$~\cite{Blundell} for all samples. We estimate $\mu_{eff}$ = 2.01 $\mu_B$/f.u. for parent compound, which appear greater than the expected Hund's rule value of 1.73 $\mu_B$/f.u. for $S = 1/2$. Similar discrepancy [i.e. obtained experimental value of $\mu_{eff}$ being larger than expected theoretical value for spin 1/2] has also been reported elsewhere~\cite{Harish,Harish2}. Such disagreement with Hund's rule value is not unusual in presence of crystal field effect and strong spin-orbit coupling. It is observed that $\mu_{eff}$ increases as $Cr$ concentration increases [Fig.~\ref{fig:mt}d]. Generally, assuming spin-only contribution for the $Cr^{3+}$($3d^3$) gives the magnetic moment of 3.9 $\mu_B$/Cr. On the other hand, calculation of the same in the strong SOC regime gives 0.33 $\mu_B$/Ir. Theoretical effective magnetic moment $\mu_{eff}$ per f.u. can be calculated as $\mu_{eff}^{theo} = (2-x)\mu_{Ir}^2 + x\mu_{Cr}^2$. The estimated $\mu_{eff}^{theo}$ turns out to be 2.57$\mu_B$/f.u., 2.68$\mu_B$/f.u. and 2.9$\mu_B$/f.u. for x = 0.05, 0.1 and 0.2 samples, respectively. The enhancement in $\mu_{eff}$ with x is anyway expected due to the substitution of the high $Cr$ moment for the low moment of $Ir$.

\begin{figure}
\includegraphics[width=8.5cm]{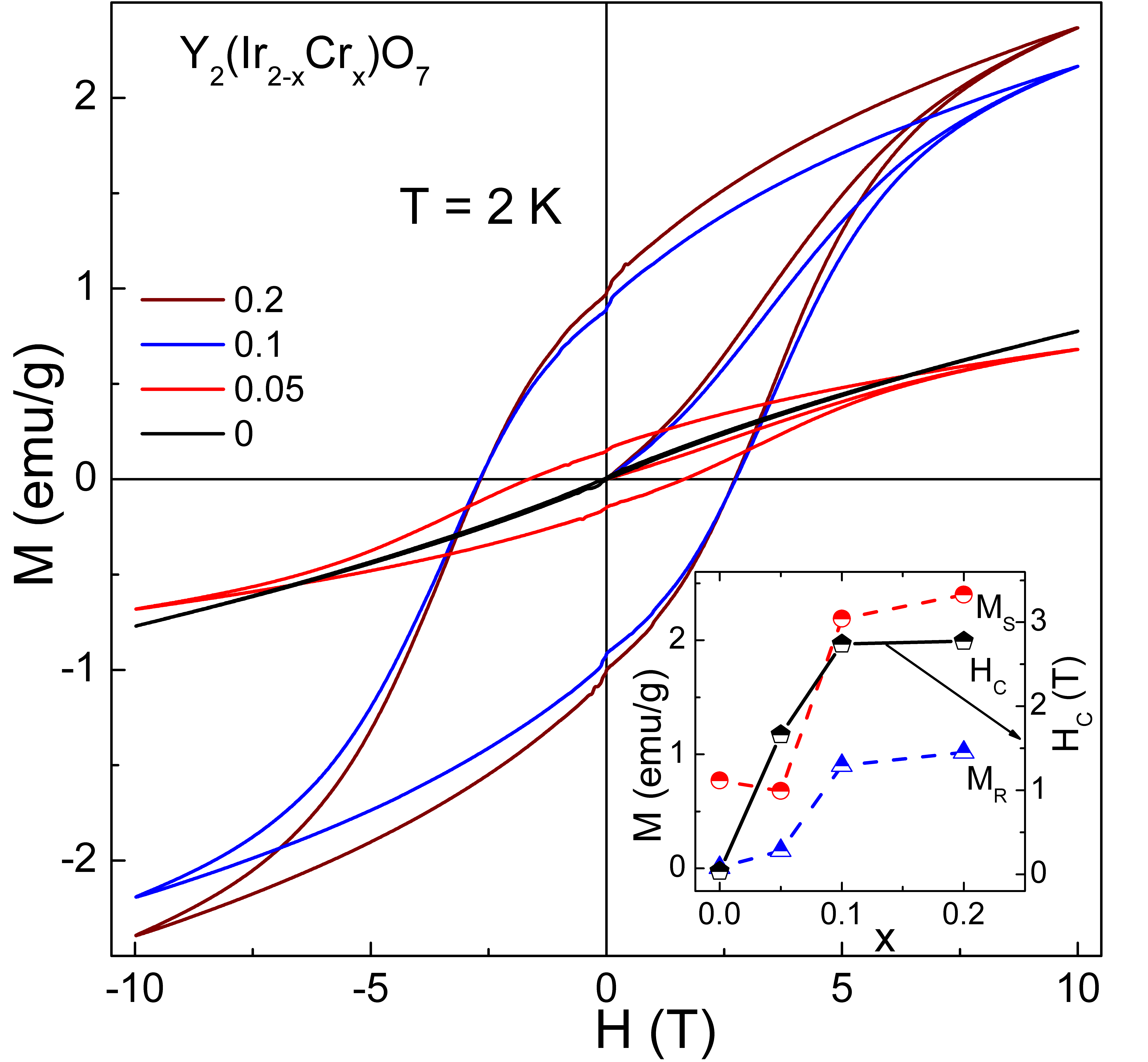}\\
\caption{Magnetic hysteresis loops (M-H) measured at temperature 2 K for $Y_2Ir_{2-x}Cr_xO_7$. Inset shows saturation magnetization ($M_S$), remanant magnetization ($M_R$) on left y-axis while coericive magnetic field ($H_C$) on right y-axis as a function of $Cr$ doping content x.}\label{fig:mh}
\end{figure}

The magnetization (M) as a function of magnetic field (H) for all samples measured at temperature 2K are shown in Fig.~\ref{fig:mh}. It is clear that $Cr$ doping leads to enhanced magnetization with clear hysteresis loop as shown in Fig.~\ref{fig:mh} suggesting enhancement of FM component vis-a-vis the AFM background. Strikingly, M-H curves of $Cr$ doped compounds shown in Fig.~\ref{fig:mh} do not show saturation up to 10T. This could be due to the coexistence of antiferromagnetic and ferromagnetic interactions, leading to magnetic frustration. As can be noticed that both $M_S$ and $M_R$ increases with increasing $Cr$ doping concentration consistent with the increase of weak ferromagnetic correlation induced by the double-exchange interaction between $Ir^{4+}$ and $Cr^{3+}$ ions. The coercive field $H_C$ also increases with $Cr$ doping shown in inset of Fig.~\ref{fig:mh}. The introduction of $Cr$ should reduce the spin-orbit coupling because of $Cr^{3+}-3d^3$ (low atomic number) replacing $Ir^{4+}-5d^5$ (high atomic number). Therefore, magnetocrystalline anisotropy is not responsible for the increase in $H_C$. Another possibility might be pinning of domain wall, which emerges from the frustration of antiferromagnetic phase induced by the randomly distributed $Cr$ ion on the $Ir$-site. Similar increase in $H_C$, $M_S$ and $M_R$ have been reported in other disordered magnets~\cite{Dho}. To summarize, cluster-glass-like characteristic with weak ferromagnetic correlation are observed in $Cr$-doped $Y_2Ir_2O_7$ samples.

\begin{figure}
\includegraphics[width=9cm]{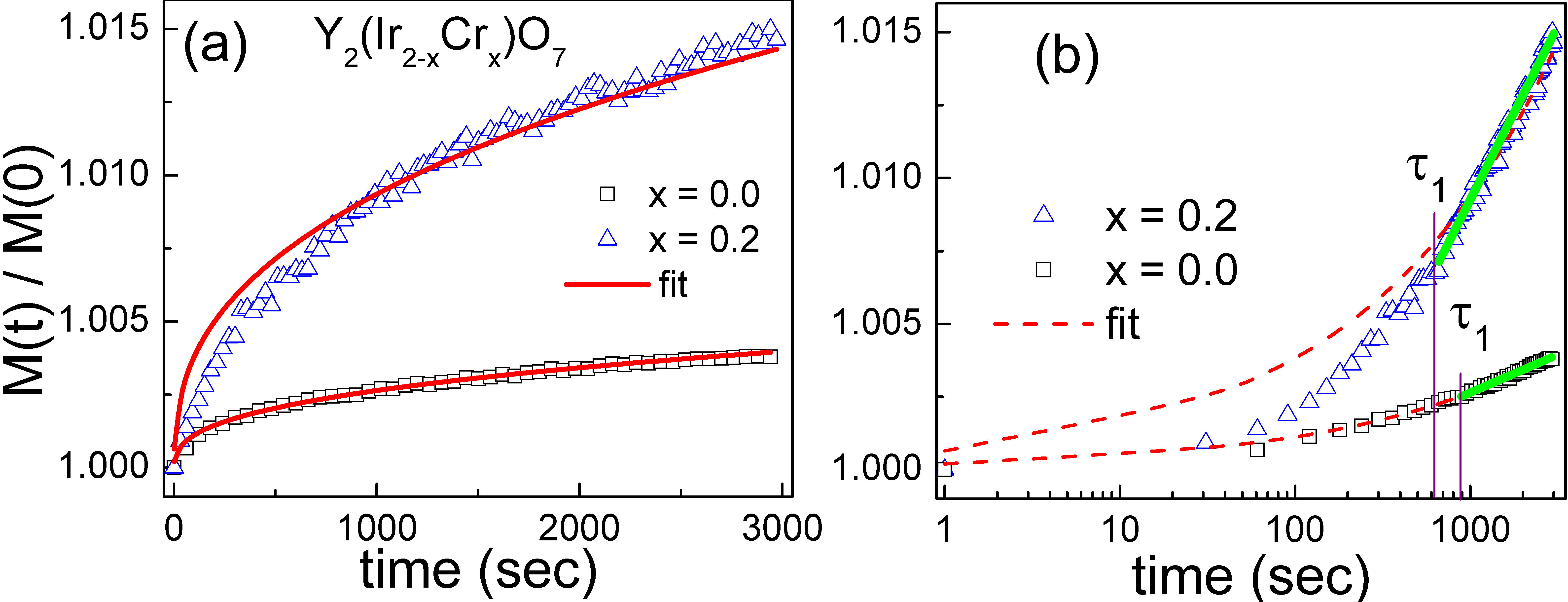}\\
\caption{(a) The normalized isothermal remanent magnetization as a function of time measured at temperature 5K with waited time $t_w$ = 10$^3$s for the x = 0.0 and 0.2 samples. Red solid line represents the fit using Eq.~\ref{eq:sg}. (b) Semi-log plot of time dependence of the normalized magnetization. Red solid line represents the fit using Eq.~\ref{eq:sg} and green line displays the linear portion.}\label{fig:Relax}
\end{figure}

In order to further confirm the glassy characteristic in $Cr$ doped compounds, the isothermal remanent magnetization is measured by cooling the sample in an applied magnetic field H = 0 from room temperature to 5K. After stabilizing the temperature and waiting upto 10$^3$s, magnetic field H = 1kOe is applied, and magnetization as a function of time is recorded. Figure~\ref{fig:Relax}a shows time dependent isothermal remanent magnetization data normalized with magnetization value M(t=0) for two representative samples x = 0.0 and 0.2. It can be seen that M(t)/M(t=0) increases with time without any sign of saturation for all the representative samples. 

 We have fitted the normalized magnetic relaxation data using stretched exponential function as shown below~\cite{Harish2,Tiwari}
\begin{equation}\label{eq:sg}
\frac{M(t)}{M(t=0)} = exp\left[\frac{t}{\tau}\right]^\beta
\end{equation}
where $\tau$ is the characteristic relaxation time, $\beta$ is the stretching exponent. The value of stretching exponential falls in the range $0 < \beta < 1$. The value $\beta$ = 1 represents the magnetic relaxation behaviour arising from a single energy barrier. On the other hand, existence of a distribution of relaxation time produces stretched exponential behaviour. The solid red lines in Fig.~\ref{fig:Relax} represent fitting of data according to Eq.~\ref{eq:sg}. The obtained $\tau$ and $\beta$ are in good agreement with corresponding values for classical spin glass systems $Ag:Mn$~\cite{Hoogerbeets}. It can be seen that doping  reduces the relaxation time $\tau$ almost by one [x = 0.2, $\tau \sim 1.7\times 10^8$] order as compared to parent compound [$\tau \sim 8.5\times 10^9$. This obviously demonstrates that doping helps the spins arrangement to relax at a faster rate. Simultaneously, there is an enhancement of the exponent $\beta \sim$ 0.39(x = 0.2) compared to parent sample ($\beta \sim$ 0.37). Figure~\ref{fig:Relax}b shows semi-logarithmic plot of normalized relaxation data as a function of time. All the samples show continuous increase of magnetization with time, do not show any sign of saturation at higher time scale. It suggests existence of uniform distribution of relaxation time of finite width, i.e. $\bigtriangleup \tau = \tau_1 - \tau_2$~\cite{Sirena}. In this regime magnetization enhances logarithmically as shown by green line in Fig.~\ref{fig:Relax}b. The lower limit $\tau_1$ is indicated in Fig.~\ref{fig:Relax}b.

\begin{figure}
\includegraphics[width=9cm]{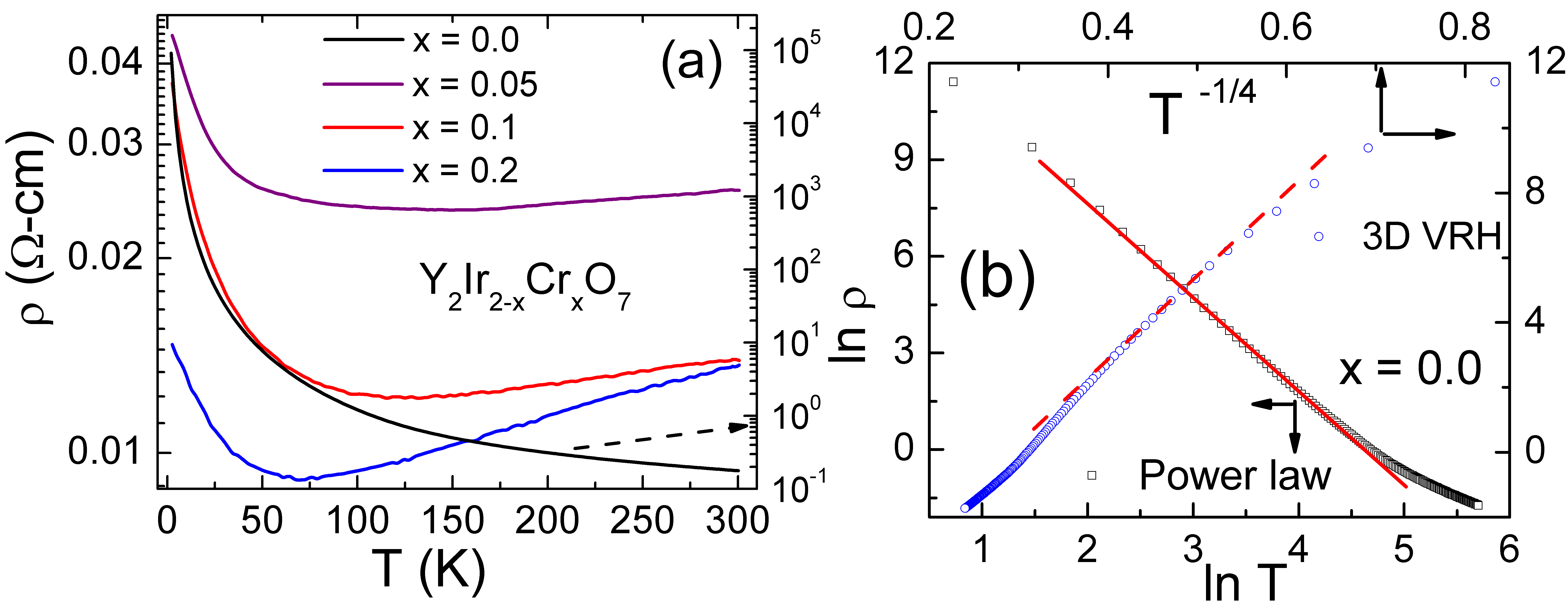}\\
\caption{(a) Semi log plot of resistivity ($\rho$) versus temperature for the samples x = 0.0 (shown on right y-axis), and 0.05, 0.1, 0.2 (shown and left y-axis) (b) $\rho(T)$ data for x = 0.0 sample: the solid red line represent power law fitting. Logarithm of $\rho(T)$ with $T^{-1/4}$ for parent compound is shown on right y-axis; The dashed red line is the least-squares fit to the 3-dimensional Mott variable-range-hopping model.}\label{fig:RT}
\end{figure}

\begin{figure*}[t]
\includegraphics[width=17.5cm]{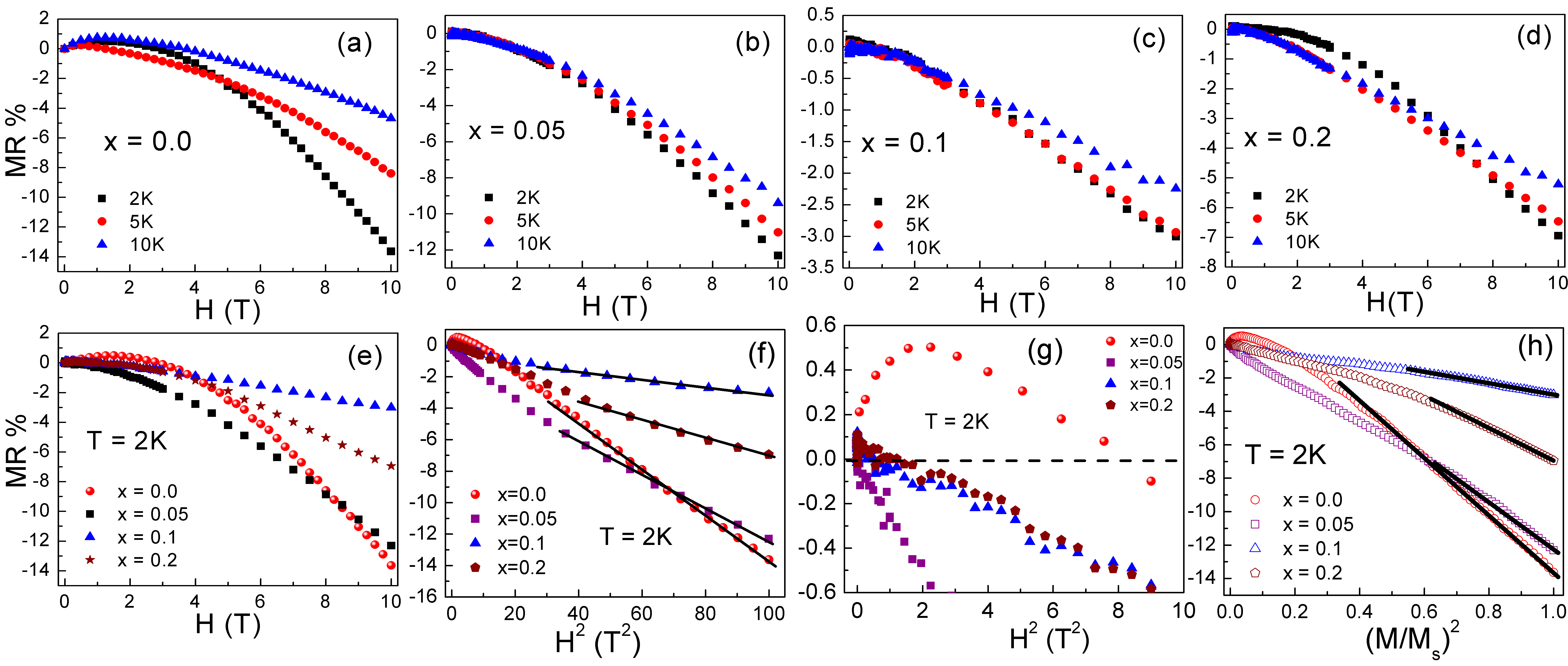}\\
\caption{Magnteoresistance (MR) as a function of magnetic field (H) measured at temperature T = 2K, 5K and 10K for (a) x = 0.0, (b) x = 0.05, (c) x = 0.1, and (d) x = 0.2 samples. (e) MR vs H of x = 0.0, 0.05, 0.1, and 0.2 samples measred at temperature 2K. (f) Quadratic magnetic field dependence of MR. (g) Enlarged view of the low field portion of MR vs H$^2$ data. (h) Linear variation of MR as a function of square of reduced magnetization.}\label{fig:MR}
\end{figure*}

Figure~\ref{fig:RT}a shows temperature dependent resistivity for the undoped, and $Cr$ doped compounds, respectively. The parent compound [(shown on right y-axis) in Fig.~\ref{fig:RT}a] shows an insulating trend throughout the temperature regime. To understand the conduction mechanism at low temperature, the $\rho(T)$ data for undoped sample is analyzed by fitting to the power law in the range 10-70K [$\rho = \rho_0T^{-n}$ where n is the power law exponent \& $\rho_0$ is the prefactor, respectively] and Mott variable range hopping (VRH) expected for a 3-dimenstional (3D) disordered system [$\rho(T) = \rho_0 exp(T_0/T)^{1/4}$, where $T_0$ is the characteristic temperature], are shown in Fig~\ref{fig:RT}b. The fitting suggests validity of power law description in the intermediate temperature range as opposed to the VRH model. Similar power law driven electronic transport has been observed for undoped compound by other groups~\cite{Vinod1}. The fitting parameters are found to be $\rho_0$ $\sim$ 7.4 $\times$ 10$^5 \Omega-cm$, n $\sim$ 2.97. The replacement of $Ir^{4+}$($5d^5$) with $Cr^{3+}$($3d^3$) doping significantly reduces the electrical resistivity leading to metal-insulator transition [Fig.~\ref{fig:RT}a]. The T$_{MI}$ decreases monotonically as ionic radius of $Ir$ site decreases. The doping of $Cr$ has two effects: $1$) The reduction in $Ir$ site ionic radius due to $Cr$ doping increases the $A$ site ionic radius, which might reduce the electrical resistivity by reducing the trigonal compression on the $IrO_6$ octahedra~\cite{Krempa,Koo}. $2$) The $Cr^{3+}$($3d^3$) and $Ir^{4+}$($5d^5$) states have similar electron filling in their $t_{2g}^{3}$ and $t_{2g}^{5}$ band, respectively with a $S=1/2$ state, effectively leading to hole doping, which could possibly increase the valence state of $Ir$ from $Ir^{4+}$ to $Ir^{5+}$. In pyrochlore iridates $A_2Ir_2O_7$ the $Ir^{4+}$ has a fully filled $J_{eff}$ = 3/2 level and an unpaired half filled $J_{eff}$ = 1/2 level which is localized due to electron-electron interaction~\cite{Wan,Pesin}. On the other hand, $Ir^{5+}$ has an empty $J_{eff}$ = 1/2 level that would promote the hopping of electrons from the nearby $Ir^{4+}$ ions, leading to the delocalization of electrons and enhancement of electrical conductivity. Later on we shall come back to this point.

 Figure~\ref{fig:MR}a,b,c,d shows the normalized magnetoresistance (MR) defined as $\left[MR\% = \frac{\rho(H)-\rho(0)}{\rho(0)}\times 100\right]$ for $Cr$ doped $Y_2Ir_2O_7$ compounds measured at temperature 2K, 5K and 10K.  All the doped compounds exhibit negative MR [i.e. resistance reduces with application of magnetic field]. On the other hand, parent sample x = 0.0 shows positive MR at low magnetic field followed by negative MR with quadratic field dependence at higher magnetic field [Fig.~\ref{fig:MR}a,e,f,g] as reported earlier by the same authors~\cite{Vinod1}. Positive MR arises due to the `weak-antilocalization' effect in the materials with strong spin-orbit coupling~\cite{Marcus}. Fig.~\ref{fig:MR}f shows the quadratic field dependent magnetoresistance (MR) recorded at temperature 2 K. It can be noticed that with $Cr$ doping at $Ir$-site, the magnitude of negative MR at high field decreases upto x = 0.1 [Fig.~\ref{fig:MR}e] although quadratic field dependence shown in Fig.~\ref{fig:MR}f is still observed which could be attributed to suppression of spin fluctuation. Additionally, the MR for all the samples scales with saturation magnetization which is a property of double exchange systems near magnetic transition as shown in Fig.~\ref{fig:MR}h. The disappearance of positive MR at low magnetic filed [which is a consequence of strong SOC] in $Cr$ doped $Y_2Ir_2O_7$ samples imply reduction of SOC due to the replacement of $Ir$ with lighter $Cr$. 

\begin{figure}
\includegraphics[width=8.5cm]{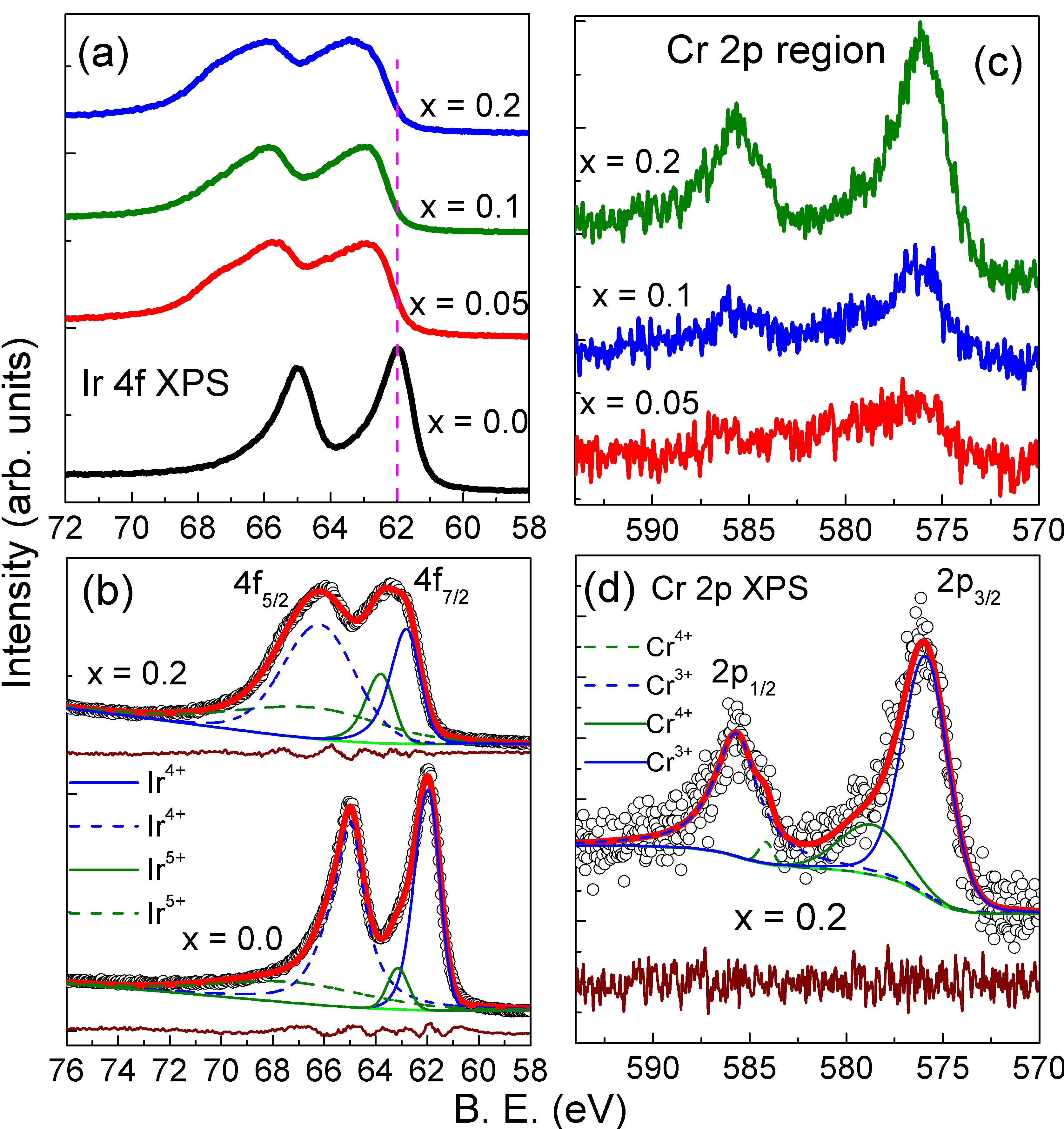}\\
\caption{High resolution XPS of (a) $Ir$ 4f for the x = 0.0, 0.05, 0.1, 0.2 samples, (b) deconvoluted $Ir$ 4f peaks for x = 0.0 and 0.2 samples, (c) $Cr$ 2p for samples x = 0.05, 0.1, 0.2, (d) fitting of $Cr$ 2p region for x = 0.2 sample.}\label{fig:XPS}
\end{figure}

Figure~\ref{fig:XPS}(a) shows normalized $Ir$ 4f XPS of the $Y_2Ir_{2-x}Cr_xO_7$ compounds. It is clear that $Cr$ doped samples exhibit more asymmetric shape than parent compound. Asymmetry in $Ir$ 4f line shapes can be attributed to the $5d$ conduction electron screening and presence of shakeup satellites above the main $Ir$ 4f line~\cite{Wertheim,Lebedev}, which could also explain the enhanced conductivity of the doped samples. Simultaneously, the less conducting bulk sample shows almost symmetrical $Ir$ 4f XPS line shape, compared to $Cr = 0.2$ doped compound. We de-convoluted the $Ir$ 4f core-level XPS spectra of x = 0.0 and 0.2 samples shown in Fig.~\ref{fig:XPS}b using asymmetric Gauss-Lorentz sum function. The observed peaks are indexd according to previous report~\cite{Vinod1}. We find that although the major contribution is due to the $Ir^{4+}$ charge states, there is a small contribution from the $Ir^{5+}$ in the parent compound. The appearance of very little amount of $Ir^{5+}$ oxidation state is consistent with what several groups have reported~\cite{Harish2,Zhu,Vinod1,Lebedev,Pei}. For $Cr$ doped samples, the contribution from $Ir^{5+}$ is enhanced, suggesting coexistence of mixed oxidation states of $Ir$, i.e. $Ir^{4+}$ and $Ir^{5+}$.

 We further analyzed the oxidation state of $Cr$ using XPS spectra of $Cr$ 2p core-level, as shown in Fig.~\ref{fig:XPS}c,d. Figure~\ref{fig:XPS}c shows the variation in the $Cr$ 2p XPS spectra with different doping concentration of $Cr$. For lowest doping content x = 0.05, it is difficult to identify the $Cr$ $2p_{3/2}$ and $Cr$ $2p_{1/2}$ peaks. While on the other hand for x = 0.1 and 0.2 samples, both peaks are clearly visible. Figure~\ref{fig:XPS}d shows the XPS spectra of $Cr$ 2p core-level, which indicates the coexistence of $Cr^{3+}$ and $Cr^{4+}$ charge~\cite{Ashish}. The XPS spectra for $Y$ 3d in x = 0.0,0.2 compounds [not shown here] suggest the existence of only $Y^{3+}$ oxidation states for undoped and $Cr$ doped samples. 

\begin{figure}
\includegraphics[width=8.5cm]{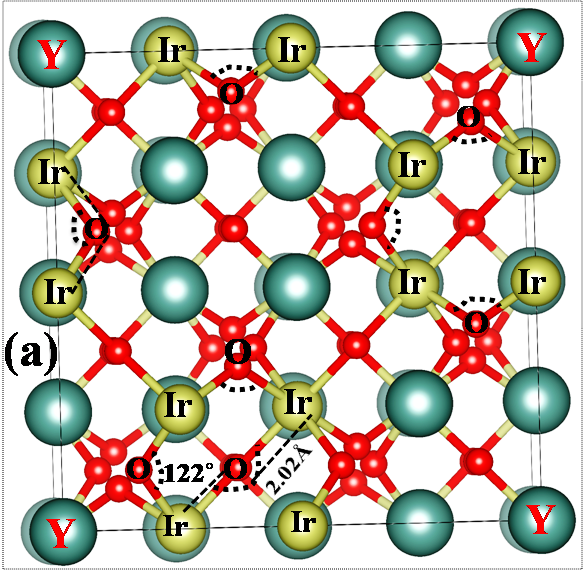}\\
\includegraphics[width=8.5cm]{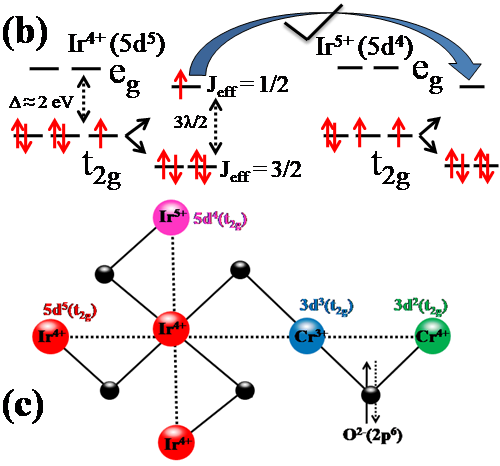}\\
\caption{(a) Unit cell structure of $Y_2Ir_2O_7$. (b) A schematic diagram of electron population in their respective orbitals demonstrate spin hopping of $J_{eff}$ = 1/2 level electron between $Ir^{4+}$ and $Ir^{5+}$ combination, represents the mechanism of enhanced ferromagnetism and electrical conductivity. (c) Possible arrangement of atoms in $Y_2Ir_{2-x}Cr_xO_7$ compound.}\label{fig:cryst}
\end{figure}

Figure~\ref{fig:cryst}a shows unit cell structure of $Y_2Ir_2O_7$ after refining the XRD data. The calculated $Ir-O-Ir$ bond angle and $Ir-O$ bond length are consistent with previous reports~\cite{Wan,Wan1}. The oxygen atom is placed at off-centered position and shared by $Ir-O_6$ octahedra in the unit cell. So far as $Y_2Ir_{2-x}Cr_xO_7$ series is concerned, $Cr^{3+}$ and $Ir^{4+}$ are magnetically active with their S=1/2 electrons. The filling of electrons in their respective $d$-orbitals are shown in Fig.~\ref{fig:cryst}b. It is known that the $IrO_6$ octahedral oxygen environment in $Y_2Ir_2O_7$ is slightly distorted due to elongation along crystallographic c-axis, leading to lifting of degeneracy with energy level splitting $\bigtriangleup\approx 2eV$. Furthermore, the octahedral environment splits $d$ orbital into $e_g$ and $t_{2g}$ orbitals. The large SOC further splits the $t_{2g}$ levels into a half filled $J_{eff}=1/2$ level with a double-degeneracy and a completely filled $J_{eff}=3/2$ level with a quadruple degeneracy. Hence, in the strong SOC dominated picture, $Ir$ gives a magnetic moment of value $0.33\mu_B/Ir$. Finally, the on-site Coulomb repulsion U further splits the $J_{eff}=1/2$ level and opens up a Mott-like gap which makes such systems $J_{eff}=1/2$ insulators. The population of $Ir^{4+}$ valence electrons with spin up (solid red arrow) and spin down in $t_{2g}$ orbital, $J_{eff}=1/2$ and $J_{eff}=3/2$ is shown in Fig.~\ref{fig:cryst}b. While on the other hand, the $3d^3$ electrons contributed by $Cr^{3+}$ populate $d_{xy}$, $d_{xz}$ and $d_{yz}$ orbitals of the $t_{2g}$ level as well, leading to $Ir^{5+}$ valence state which should be non-magnetic considering four electrons will fully occupy the $J_{eff}=3/2$ quartet state.

Now let us recall the major consequences of the doping of $Cr^{3+}$ at $Ir^{4+}$ site in the $Y_2Ir_2O_7$ compound on the magnetic and electronic properties: 1) The AFM correlations weaken against $Cr$ doping. 2) Electronic conductivity is enhanced with doping. In $A_2Ir_2O_7$ pyrochlore systems, magnetic interactions take place at the corner-shared $Ir-O_6$ octahedra through the mediation of $O-2p$ orbital. Based on the present scenario, with random ionic distribution, a schematic representation of the magnetic interaction for $Y_2Ir_{2-x}Cr_xO_7$ series is shown in Fig.~\ref{fig:cryst}c. We propose that in $Y_2Ir_{2-x}Cr_xO_7$ compounds exchange interaction takes place primarily through the $Ir^{4+}-O^{2-}-Ir^{4+}$, since $Ir^{5+}$ is non-magnetic. With $Cr$ doping, the reduction in the lattice parameter, $Ir-O-Ir$ bond angle and enhancement in the $p$ value, $Ir-O$ bond length are shown in Fig.~\ref{fig:bond}a,b,c,d. In such a situation, the $Ir$ atoms reside on a more distorted $Ir-O_6$ octahedra than parent compound as $Ir-O-Ir$ bond angle tends towards $90^0$ which might be responsible for weakening of AFM correlations and higher value of magnetic moment. The distorted $Ir-O_6$ octahedra along the c-axis could weaken AFM correlation, as indicated by reduction in absolute value of Curie-Weiss temperature $\theta_{CW}$.

\section{Conclusion}
We have investigated the structural, magnetic, and electronic properties of the pyrochlore iridates $Y_2Ir_{2-x}Cr_xO_7$. The introduction of $Cr^{3+}$ at $Ir^{4+}$ sites weakens the antiferromagnetic correlation and enhances electrical conductivity. The XRD analysis shows distorted $Ir-O_6$ octahedra and reduction in $Ir-O-Ir$ bond angle as $Cr^{3+}$ doping concentration increases. The X-ray photo-emission spectroscopy measurements suggest the coexistence of $Ir^{4+}$ and $Ir^{5+}$ in the $Y_2Ir_{2-x}Cr_xO_7$ compounds, where the amount of $Ir^{5+}$ enhances with $Cr$ doping. This explains the possible origin of the weak ferromagnetism and enhanced electrical conductivity in the same. The cusp in ZFC M-T curves, the irreversiblity in FC-ZFC magnetization at higher temperature and hysteretic isothermal magnetization with large coercive field for $Y_2Ir_{2-x}Cr_xO_7$ at low temperature suggest cluster-glass like transition rather than long range ferromagnetic ordering. This is also confirmed by relaxation measurements. We emphasize that doping affects the local chemistry such as bond angle, bond length and oxidation states which in turn influences, \textit{in not neccessarily inter-connected fashion}, the electronic transport and magnetic properties in 5d iridates.

\end{document}